\documentstyle[prl,floats,aps]{revtex}

\begin{document}

\input{epsf.sty}

\draft

\twocolumn[\hsize\textwidth\columnwidth\hsize\csname
@twocolumnfalse\endcsname

\title{Isometric embeddings of black hole horizons in
  three-dimensional flat space}

\author{
Mihai Bondarescu${}^{(1,2)}$,
Miguel Alcubierre${}^{(1)}$,
Edward Seidel${}^{(1,3)}$ \medskip
}

\address{
$^{(1)}$ Max-Planck-Institut f\"ur Gravitationsphysik,
Albert-Einstein-Institut, Am M\"{u}hlenberg 1, 14476 Golm, Germany
}

\address{
$^{(2)}$ California Institute Of Technology,
1200 E California Blvd, Pasadena, CA 91125
}

\address{
$^{(3)}$ National Center for Supercomputing Applications,
Beckman Institute, 405 N. Mathews Ave., Urbana, IL 61801
}

\date{\today; AEI-2001-118}

\maketitle


\begin{abstract}
  The geometry of a two-dimensional surface in a curved space can be
  most easily visualized by using an isometric embedding in flat
  three-dimensional space.  Here we present a new method for embedding
  surfaces with spherical topology in flat space when such a embedding
  exists. Our method is based on expanding the surface in spherical
  harmonics and minimizing for the differences between the metric on
  the original surface and the metric on the trial surface in the
  space of the expansion coefficients.  We have applied this method to
  study the geometry of back hole horizons in the presence of strong,
  non-axisymmetric, gravitational waves (Brill waves).  We have
  noticed that, in many cases, although the metric of the horizon
  seems to have large deviations from axisymmetry, the intrinsic
  geometry of the horizon is almost axisymmetric.  The origin of the
  large apparent non-axisymmetry of the metric is the deformation of
  the coordinate system in which the metric was computed.
\end{abstract}

\pacs{04.25.Dm, 04.30.Db, 95.30.Sf, 97.60.Lf}

\narrowtext

\vskip2pc]


\section{Introduction}
\label{sec:intro}

In the past few years, important progress has been made in the
development of fully three-dimensional (3D) numerical relativity
codes.  These codes are capable of simulating the evolution of
strongly gravitating systems, such as colliding black holes and neutron
stars, and can provide important physical information about those
systems such as the gravitational waves they produce.  The codes also
allow one to locate and track the evolution of apparent and event
horizons of black holes that might exist already in the initial data
or might form during the evolution of the spacetime.  However, since
the location of such horizons is obtained only in coordinate space,
one typically has little information about the real geometry of those
surfaces.  One can, for example, obtain very similar shapes in
coordinate space for horizons that are in fact very different (see for
example the family of distorted black holes studied
in~\cite{Anninos93a}; their coordinate locations are very similar, but
their geometries are quite different). The most natural way to
visualize the geometry of a black hole horizon, or of any other
surface computed in some abstract curved space is to find a surface in
ordinary flat space that has the same intrinsic geometry as the
original surface.  The procedure of finding such a surface is called
{\em embedding}\/ the surface in flat space.

It is a well known fact~\cite{Berger98} that any two dimensional
surface is {\em locally}\/ embeddable in flat 3D space.  Global
embeddings of a surface, on the other hand, might easily not exist,
and even when they do they are not easy to find.  However, several
methods have been proposed in the past for computing partial (or
global) embeddings of surfaces when such embeddings exist.

Partial embeddings of a slice through the Misner initial data for
colliding black holes~\cite{Misner60} have been computed for example
in~\cite{Romano95}. The method used to find such embeddings starts
from the metric of the original surface written in terms of
some local coordinates $(u,v)$:
\begin{equation}
ds^2 = E du^2 + 2F du dv + G dv^2 \, .
\end{equation}
One then introduces embedding functions $X(u,v)$, $Y(u,v)$ and
$Z(u,v)$ such that
\begin{equation}
dX^2 + dY^2 + dZ^2 = ds^2 = E du^2 + 2F du dv + G dv^2 \, ,
\end{equation}
which implies
\begin{eqnarray}
E &=& X_{,u}^2 + Y_{,u}^2 + Z_{,u}^2 \, ,
\label{eq:Romano_E} \\
F &=& X_{,u} X_{,v} + Y_{,u} Y_{,v} + Z_{,u} Z_{,v} \, ,
\label{eq:Romano_F} \\
G &=& X_{,v}^2 + Y_{,v}^2 + Z_{,v}^2 \, .
\label{eq:Romano_G} 
\end{eqnarray} 

The above system of non-linear first order partial differential
equations is not of any standard type. In order to solve it one can
use a method originally proposed by Darboux. This leads to a single
non-linear second order partial differential equation for $Z(u,v)$
known as the Darboux equation.  The character of the Darboux equation
depends on the sign of the Gaussian curvature $K$ of the surface and
on the orientation of the embedding at the point of integration as
follows: for $K \geq 0$ the equation is elliptic, for $K \leq 0$ it is
hyperbolic and it has parabolic character both if $K=0$ or if the
surface is vertical.  For the Misner geometry, the curvature is always
negative and the Darboux equation is hyperbolic.  It can then be
rewritten by using the characteristics as coordinates, and solved as a
Cauchy problem given appropriate initial data.  Once the Darboux
equation has been solved for $Z(u,v)$, the remaining equations can be
integrated to give $X(u,v)$ and $Y(u,v)$.

This method is not appropriate for embedding surfaces that have both
regions with $K>0$ and regions with $K<0$.  The reason for this is
that the Darboux equation one has to integrate is elliptic for $K>0$
and hyperbolic for $K<0$.  This means that for such surfaces the
Darboux equation will change type and its integration will become very
difficult.  This will typically be the case for surfaces with
spherical topology, such as black hole horizons, which will always
have some regions of positive curvature, and may well have regions of
negative curvature too.

One of the first studies of the intrinsic geometry of rotating black
hole horizon surfaces was carried out by Smarr~\cite{Smarr73b}.  There
it was show, by direct construction of the embedding from the analytic
Kerr metric, that while the horizon of a Schwarzschild black hole is
spherical, for rotating black holes the horizon has an equatorial
bulge, a satisfying and intuitive result that reinforces the notion
that geometric studies of black hole horizons can add physical
insight.  The equatorial bulge can be characterized by an oblateness
factor that is uniquely determined by the ratio $a/m$, where $m$ is
the mass of the black hole and $a$ its rotation parameter.  It was
also shown that for rapidly rotating black holes, with $a/m >
\sqrt{3}/2$, the Gaussian curvature becomes negative near the poles,
and the surface is not embeddable in Euclidean space, as it is ``too
flat''!

Extending on this work, using a direct constructive embedding method
described below, a number of studies were made of distorted, rotating,
and colliding black hole horizons in
axisymmetry~\cite{Bernstein93a,Anninos93a,Anninos93d,Anninos94f,Anninos95b,Anninos95c}
where it was shown that embeddings are very useful tools to aid in the
understanding of the {\em dynamics} of black holes.  For example,
distorted rotating black hole horizons were found to oscillate, about
their oblate equilibrium shape, at their quasi-normal frequency.  The
recent work on isolated
horizons~\cite{Ashtekar98a,Ashtekar99a,Ashtekar00a,Ashtekar01a} shows
how geometric measurements of the horizon can be used to determine,
for example, the spin of a black hole formed in some process, and
other physical features.

However, in the absence of axisymmetry, the problem of constructing an
embedding for a black hole horizon becomes much more difficult.  One
approach to compute such embeddings of horizons in 3D spacetimes has
been suggested by H.-P. Nollert and H. Herold~\cite{Nollert98}.  They
consider a triangular wire frame on the original surface and compute
the distances between each point and its neighbors using the intrinsic
metric of the surface.  They then consider a network with the same
topology in flat space and try to solve the system of equations
\begin{equation}
|{\bf r}_i-{\bf r}_j| = d_{ij} \, ,
\end{equation}
where ${\bf r}_k$ represents the position vector of the
$k^{\text{th}}$ point in flat space and $d_{ij}$ represents the
distance between the points $P_i$ and $P_j$ computed on the original
surface.  If necessary, they refine the grid until they reach a
desired accuracy.

The approach of Nollert and Herold seems very natural, but is has the
serious drawback that it does not always converge to the correct
solution.  The reason for this is that the method imposes constraints
only on the distances between points, but it does not guarantee that the
final surface will be smooth. There are in fact multiple solutions to
the system of equations, and for most such solutions the resulting
embedding is not smooth.  For example, if one tries to embed a simple
sphere, this method might indeed converge to the sphere, but it might
also converge to the surface one obtains if one cuts the top of the
sphere, turns its up-side down and glues it back.  The distances
between point are the same in both surfaces, but only one of them is
smooth.

The method for computing embeddings that we present in this paper is
based on a spectral decomposition of the surface in spherical
harmonics written in a non-trivial mapping of the coordinate system.
We search for the embedding by minimizing for the difference between
the metric of the original surface and that of our trial embedding in
the space of the coefficients of the spherical harmonics and of the
coordinate mappings.  Since we use a decomposition of the surface
in spherical harmonics, the surface is guaranteed to be smooth.  By
increasing the number of spherical harmonics used in the decomposition
of the physical surface, one can get as close to the correct embedding
as desired.


\section{Our Method}

The intrinsic geometry of any surface is completely determined by its
metric.  To construct the embedding of a given surface $S$, one needs
to find a surface $S'$ in flat space that has the same metric as $S$
in an appropriate coordinate system.  It is important to stress here
the fact that finding the embedding surface $S'$ also requires that
one finds an appropriate mapping of the original coordinate system in
the surface $S$ to a new coordinate system in the surface $S'$ in
which the two metrics are supposed to agree.

\subsection{A Direct Method for Horizon Embeddings in Axisymmetry}

Before describing our new method, it is instructive to describe a
simple and direct method for axisymmetric embeddings of horizons, used
by a number of authors to study the physics of dynamic black
holes~\cite{Bernstein93b,Anninos93a,Anninos93d,Anninos94f,Anninos95b,Anninos95c,Masso98c}.

For the case of non-rotating, axisymmetric spacetimes (easily
generalized to rotation, but restricted here merely for ease of
illustration), the 3D metric on a given time slice can be written as
\begin{equation}
    ds_{(3)}^{2}= A(\eta,\theta)d\eta^{2} + B(\eta,\theta)d\theta^{2} + 
    D(\eta,\theta) \sin^{2}\theta d\phi^{2}.
\end{equation}
The location of the axisymmetric horizon surface is given by the
function $\eta = \eta_{s}(\theta)$.  The 2D metric induced on the
horizon surface is then given by
\begin{equation}
\label{curvedmetric}
ds_{(2)}^{2} = \left[ B + \left( \frac{d \eta_{s}}{d \theta} \right)^{2} A
\right] d\theta^{2} + D \sin^{2}\theta d\phi^{2} \, .
\end{equation}

Now, the flat metric in cylindrical coordinates $(z,\rho,\psi)$ can be
written as
\begin{equation}
    \label{flatmetric}
    ds^{2}=dz^{2}+d\rho^{2}+\rho^{2}d\psi^{2}.
\end{equation}
To create an embedding in a 3D Euclidean space, we want to construct
functions $z(\theta,\phi), \rho(\theta,\phi)$, and $\psi(\theta,\phi)$
such that we can identify the line elements given by
Eqs.~(\ref{curvedmetric}) and~(\ref{flatmetric}), that is, that all
lengths be preserved.

Here, we are faced with our first choice about the coordinates used in
the embedding, a problem which will be more complex in the general
case as we show below.  Since the spacetime itself is axisymmetric, it
is a natural choice to make the embedding axisymmetric.  We choose,
then, to construct a surface for a constant value of $\phi=0$, and we
then have $z=z(\theta), \rho=\rho(\theta)$, and the resulting
embedding will be a surface of revolution about the $z-$axis. Using
the obvious mapping between $\psi$ and $\phi$, $\psi = \phi$, it
becomes straightforward to derive ordinary differential equations to
integrate for $\rho(\theta)$ and $z(\theta)$ along the horizon
surface.  It is important to emphasize that we have to make a choice
about the embedding coordinates, even in this simpler case, as we must
in the general case discussed below.

Using this method, embeddings were carried out during the numerical
evolution for a variety of dynamic, axisymmetric black hole
spacetimes~\cite{Bernstein93b,Anninos93a,Anninos93d,Anninos94f,Anninos95b,Anninos95c,Masso98c}.
The evolution of these embeddings were found to be extremely useful
in understanding the physics of these systems.  We will use some of
these results as test cases for the more general method for 3D
spacetimes, as detailed in the next section.

\subsection{Our General Method for Embeddings in full 3D}  

Given a coordinate system $\xi^i$ ($i,j=1,2)$ on our surface, the
first step in looking for an embedding is to find the two dimensional
metric $g_{ij}$ of the surface induced by the metric of the three
dimensional space $h_{ab}$ ($a,b=1,2,3$) in which it is defined.  The
general procedure to find such induced metric is to construct a
coordinate basis of tangent vectors ${\bf e}_i := \partial_i$ on the
surface.  The induced metric will then be given by
\begin{equation}
g_{ij}(\theta,\phi) = {\bf e}_i(\theta,\phi) \cdot {\bf e}_j(\theta,\phi)
= h_{ab} \, e_i^a  \, e_j^b  \, ,
\label{eq:metric}
\end{equation}
where $e_i^a$ is the component of the vector ${\bf e}_i$ with
respect to the three dimensional coordinate $x^a$.

Since the surfaces we are concerned with in this paper (black hole
horizons) have spherical topology, we will assume that the
three-dimensional metric $h_{ab}$ is given in terms of spherical
coordinates $(r,\theta,\phi)$ defined with respect to some origin
enclosed by the surface.  Furthermore, we will also assume that the
surface is a ``ray-body'' (Minkowski's {\em
  strahlkorper}~\cite{Schroeder86}, also known as a ``star-shaped''
region) that is, a surface such that any ray coming from the origin
intersects the surface at only one point.  Such a property implies
that we can choose as a natural coordinate system on the surface
simply the angular coordinates $(\theta,\phi)$.

If we take our surface to be defined by the function
\mbox{$r=f(\theta,\phi)$}, then it is not difficult to show that the
induced metric $g_{ij}$ on the surface will be given in terms of the
three-dimensional metric $h_{ab}$ as:
\begin{eqnarray}
g_{\theta \theta} &=& h_{\theta \theta}
+ h_{rr} \left( \partial_\theta f \right)^2
+ 2 h_{r \theta} \partial_\theta f \, , \\
g_{\phi \phi} &=& h_{\phi \phi}
+ h_{rr} \left( \partial_\phi f \right)^2
+ 2 h_{r \phi}\partial_\phi f \, , \\
g_{\theta \phi} &=& h_{\theta \phi}
+ h_{rr} \, \partial_\theta f \partial_\phi f
+ h_{r \phi} \partial_\theta f 
+ h_{r \theta} \partial_\phi f\, .
\label{eq:2Dfrom3D}
\end{eqnarray}

Let us now for moment a assume that an embedding of our surface in
flat space exists, and let us also introduce a spherical coordinate
system $(r_e,\theta_e,\phi_e)$ in flat space.  Notice that there is no
reason to assume that a point with coordinates $(\theta,\phi)$ in
the original surface will be mapped to a point with the same angular
coordinates in the embedding.  In general, the embedded surface in
flat space will be described by the relations
\begin{equation}
r_e = r_e(\theta,\phi) \, , \quad \theta_e = \theta_e(\theta,\phi) \, ,
\quad \phi_e = \phi_e(\theta,\phi) \, .
\label{eq:mapping}
\end{equation}

A crucial observation at this point is that the angular coordinates
$\{\theta,\phi\}$ in the original surface still provide us with a {\em
well behaved coordinate system in the embedded surface}, only one that
does not correspond directly to the flat space angular coordinates
$\{\theta_e,\phi_e\}$, but is instead related to them by the
coordinate transformations $\theta_e = \theta_e(\theta,\phi)$ and
$\phi_e = \phi_e(\theta,\phi)$.  This means that under the embedding,
points with coordinates $(\theta,\phi)$ in the original surface will
be mapped to points with {\em the same} coordinates $(\theta,\phi)$ in
the embedded surface, but {\em different} coordinates
$(\theta_e,\phi_e)$.  We then have two natural sets of coordinates in
the embedded surface: the ones inherited directly from the original
surface through the embedding mapping, and the standard angular
coordinates in flat space.

By definition, an embedding preserves distances, so the proper distance
between two points in the original surface must be equal to the
distance between the two corresponding points in the embedded surface.
Since those corresponding points have precisely the same coordinates
$\{\theta,\phi\}$, we must conclude that for the embedding to be
correct, the components of the metric tensor in both surfaces when
expressed in terms of the coordinates $\{\theta,\phi\}$ must be
identical.  That is, if we call $g^e_{ij}$ the metric of the embedded
surface, we must have
\begin{equation}
g_{\theta \theta} = g^e_{\theta \theta} \, , \quad
g_{\theta \phi} = g^e_{\theta \phi} \, , \quad
g_{\phi \phi} = g^e_{\phi \phi} \, .
\label{eq:equalmetrics}
\end{equation}
It is important to stress that the components of the metric in the
embedded surface will only be equal to the components of the metric in
the original surface if we use the inherited coordinate system, but
not if we use the standard angular coordinates in flat space.

Computing the components of the metric in the embedded surface in terms
of the inherited coordinate system $\{\theta,\phi\}$, given the
embedding relations~(\ref{eq:mapping}), is not difficult.  All one
needs to do in practice is consider four points in the original
surface with coordinates $P_1(\theta,\phi)$, $P_2=(\theta+\delta
\theta,\phi)$, $P_3=(\theta,\phi+\delta \phi)$, $P_4=(\theta+\delta
\theta,\phi+\delta \phi)$, find their corresponding coordinates
$\{r_e,\theta_e,\phi_e\}$ in flat space using~(\ref{eq:mapping}),
compute their squared distances using the flat space metric, and then
solve for the metric components from
\begin{eqnarray}
(\overline{P_1 P_2})^2 &=& g^e_{\theta \theta} \, d \theta^2 \, , \\
(\overline{P_1 P_3})^2 &=& g^e_{\phi \phi} \, d \phi^2 \, , \\
(\overline{P_1 P_4})^2 &=& g^e_{\theta \theta} \, d \theta^2
+ g^e_{\phi \phi} \, d \phi^2 + 2 g^e_{\theta \phi} \, d \theta d \phi \, .
\end{eqnarray}

Finding the embedding now means finding a mapping~(\ref{eq:mapping})
such that equations~(\ref{eq:equalmetrics}) are satisfied everywhere.

Let us consider first the relation between the $(\theta,\phi)$
coordinates on the original surface and the angular coordinates
$(\theta_e,\phi_e)$ in flat space.  Even if these two sets of
coordinates are not equal, we can safely assume that there is a
one-to-one correspondence between them.  Moreover, both are sets of
angular coordinates, so they have the same
behavior: $\theta$ and $\theta_e$ go from 0 to $\pi$, and $\phi$ and
$\phi_e$ go from 0 to $2\pi$ and are periodic.  From these
properties, it is not difficult to see that the most general
functional relation between both sets has the form

\begin{eqnarray}
\theta_e(\theta,\phi)&=& \theta + \sum_{n=0}^{\infty}b_{n0} \sin(n \theta)
\nonumber \\
&& + \sum_{n=1}^{\infty} \sum_{m=1}^{\infty} b_{nm} \sin(n\theta) \,
\sin(m\phi) \, ,
\label{eq:remap-theta} \\
\phi_e(\theta, \phi) &=& \phi + c_{00} \, \theta + \sum_{n=1}^{\infty}c_{n0}
\sin(n \theta) + \sum_{m=1}^{\infty}c_{0m}\sin{m\phi}
\nonumber \\
&& + \sum_{n=1}^{\infty} \sum_{m=1}^{\infty} c_{nm} \sin(n\theta) \,
\sin(m\phi) \, .
\label{eq:remap-phi}
\end{eqnarray}

The second term in the expansion for $\theta_e$ represents a general
axisymmetric re-mapping of $\theta$, while the third term
is required if axisymmetry is not assumed.  In the expression for
$\phi_e$, the second and third terms represent a possible rigid twist
of the coordinate system, and the last two terms stand for a general
dependence of $\phi_e$ on both $\theta$ and $\phi$.

For the radial coordinate $r_e$, it is also not difficult to see that
one can use a simple expansion in spherical harmonics of the form
\begin{equation}
r_e(\theta,\phi) = \sum_{l=0}^{\infty} \sum_{m=-l}^{l}
\sqrt{4\pi} \, a_{lm} Y_{lm}(\theta,\phi) \, ,
\label{eq:harmonics}
\end{equation}
where the overall normalization factor of $\sqrt{4\pi}$ has been
inserted so that $a_{00}$ is the average radius of the surface,
$a_{10}$ is its average displacement in the $z$-direction, and so
on. We will also use a real basis of spherical harmonics, for which
$m$ and $-m$ stand for an angular dependence $\cos(m\phi)$ and
$\sin(m\phi)$, instead of the complex $\exp(im\phi)$ and
$\exp(-im\phi)$.

The metric of the embedded surface will now be completely determined by
the set of coefficients $a_{lm}$, $b_{nm}$ and $c_{nm}$. The space of
these coefficients can be regarded as a vector space $V$, with any
given point in $V$ representing a surface in flat space together with
a certain coordinate mapping.

Consider now an {\em embeddable\/} surface $S$ in some arbitrary
curved space.  It is not difficult to find a real valued function $F$
defined on $V$ that has a global minimum at the point $P \in V$ for
which the metric of the embedded surface $S^e$ is the same as the
metric of the original surface. One such function is
\begin{eqnarray}
 F = \int_{\theta=0}^{\pi} \int_{\phi=0}^{2\pi} && \left[ \,
\left( g_{\theta \theta}(\theta,\phi)-g^e_{\theta \theta}(\theta,\phi) \right)^2 
\right. \nonumber \\
&&+ \left( g_{\phi \phi}(\theta,\phi)-g^e_{\phi \phi}(\theta,\phi) \right)^2
\nonumber \\
&& \left. + \left( g_{\theta \phi}(\theta,\phi)-g^e_{\theta \phi}(\theta,\phi) \right)^2
\right] \; d \theta d \phi \, .
\label{eq:horizonfunction}
\end{eqnarray}

\noindent We call the function $F$ above the ``embedding'' function.
It is easy to see that $F \geq 0$ on any point in $V$, and that $F=0$
if and only if $g=g^e$ for all $(\theta,\phi)$.  The embedding then
corresponds to the absolute minimum of $F$ in $V$. There are many
different numerical algorithms for finding minima of general functions
in multidimensional spaces.  In our code we have used Powell's
minimization algorithm~\cite{Press86}, but we are aware that other
methods might perform better.  It is important to mention that the
definition of the embedding function $F$ above is by no means unique.
Many different forms for $F$ can be constructed, in particular, one
could take into account the fact that not all metric functions have
similar magnitudes and construct an embedding function that normalizes
each term in the above expression.

One of the problems with minimization algorithms in general is that
they cannot distinguish between a global minimum (what we really want)
and a local minimum. In our case, the value of $F$ at the absolute
minimum is zero, so we can easily distinguish between a real embedding
and a wrong solution that might appear if the minimization algorithm
gets stuck in a local minimum.  However, steering the algorithm
toward the global minimum is non-trivial.  From experience we have
seen that local minima for which $F\neq 0$ do exist for our
problem.  In order to avoid them, we have found it necessary to run
first the minimization algorithm with a small number of coefficients
$a$, $b$, and $c$, and then increase the number of coefficients one by
one until we find a good solution.  This method is certainly time
consuming, but it seems to work well in the examples we have
considered so far.

Of course, in order to find a ``perfect'' embedding, one would have to
push the number of coefficients all the way to infinity. This is
numerically impossible, so in practice we just set up a given
tolerance in the value of the function $F$ and increase the number of
coefficients until we achieve that tolerance.  We also check that the
value of $F$ goes to zero exponentially as we increase the total
number of coefficients. We have seen that $n=l\sim 14$ is enough for
relatively simple surfaces like most black hole horizons.  If one
wants to embed something more complicated (like a human face, for
example) starting from its metric in some coordinate system, this
value would clearly be too small.  Another way to see whether an
embedding is good or not is to compare directly the metric of the
original surface with the metric of the resulting embedding.  If the
fit is good enough, the code has converged to the correct embedding.

One important test we have used for our algorithm is a direct
comparison of the results obtained with our code with embeddings
computed with a different code in the special case when the surface is
axisymmetric.


\section{Tests}


\subsection{Recovering a known surface}

A very simple test for our algorithm is to look for the embedding of a
surface that is known to be embeddable and has a known embedding.  In
order to do this we first construct a surface in flat space by
choosing some arbitrary values of the spherical harmonic coefficients.
We then compute the metric of this surface in the standard
$(\theta,\phi)$ angular coordinates, and give this metric as input to
our code.  The code must then recover the correct values of the
spherical harmonic coefficients {\em plus}\/ a trivial mapping of the
angular coordinates.

\begin{figure}
\epsfxsize=3.4in \epsfysize=3.4in \epsfbox{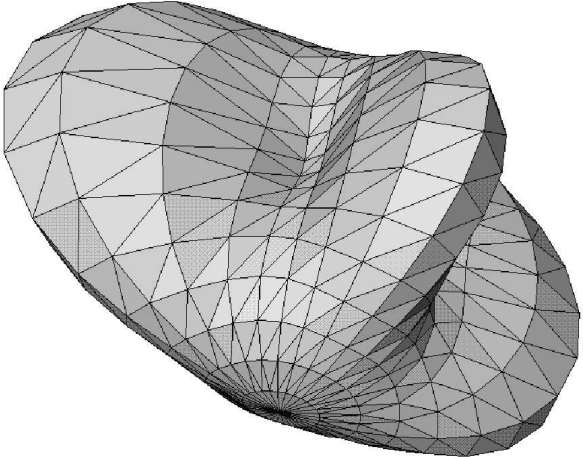}
\epsfxsize=3.4in \epsfysize=3.4in \epsfbox{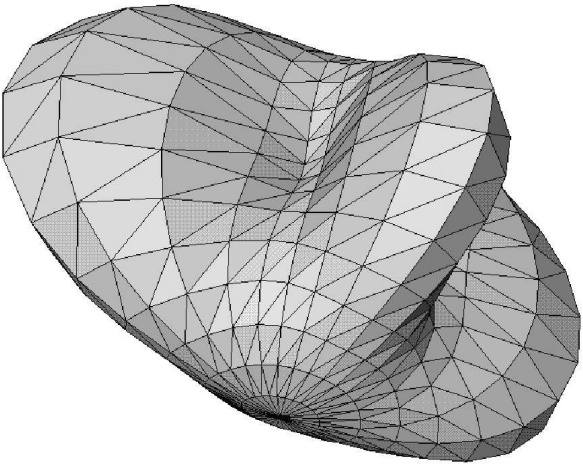}
\caption{Embedding of a test surface defined by the
  spherical harmonic coefficients ($a_{00}=9$,$a_{22}=1$,$a_{44}=4$).
  The upper panel shows the original surface and the lower panel the
  resulting embedding.}
\label{fig:handmade}
\end{figure}

We show an example of this in Fig.~\ref{fig:handmade}, where we have
chosen a surface defined by the spherical harmonic coefficients:
\begin{equation}
a_{00} = 9 \, , \quad a_{22} = 1 \, , \quad a_{44} = 4 \, ,
\label{eq:handmade}
\end{equation}
with all other coefficients equal to zero.  In the upper panel of the
figure we show the original surface, and in the lower panel the
resulting embedding. The differences in the shape of the two surfaces
are very difficult to see.

\begin{table}
\begin{tabular}{ccc}
Expansion     & Original & Recovered   \\
coefficient   & value    & value \\ \hline
$a_{00}$      &  9       & $9+1.1 \times 10^{-5} $\\
$a_{20}$      &  0       & $-3.6 \times 10^{-5}  $\\
$a_{22}$      &  1       & $1+4 \times 10^{-6}   $\\
$a_{40}$      &  0       & $1.5 \times 10^{-5}   $\\
$a_{42}$      &  0       & $-2.24 \times 10^{-5} $\\
$a_{44}$      &  2       & $2+1.9 \times 10^{-5} $
\end{tabular}
\vspace{0.2in}
\caption{Comparison of the recovered expansion coefficients for the
embedding of the test surface described in the text.}
\label{tab:coefficients}
\end{table}

For this test we have used $100 \times 100$ grid points to describe
the surface.  Since the surface is symmetric with respect to
reflections on all three $(x,y,z)$ coordinate planes, we have
considered only one octant, so the angular resolution was
\mbox{$\Delta \theta = \Delta \phi = \pi/200$}.  The recovered
expansion coefficients are shown in Table~\ref{tab:coefficients}.
The coefficients corresponding to the mapping of the angular
coordinates, as well as the rest of the $Y_{lm}$ coefficients were
either exactly zero because of the octant symmetry or had values
smaller than $10^{-3}$.

\begin{figure}
\begin{minipage}{1.6in}
\epsfxsize=1.6in
\epsfysize=3.4in
\epsfbox{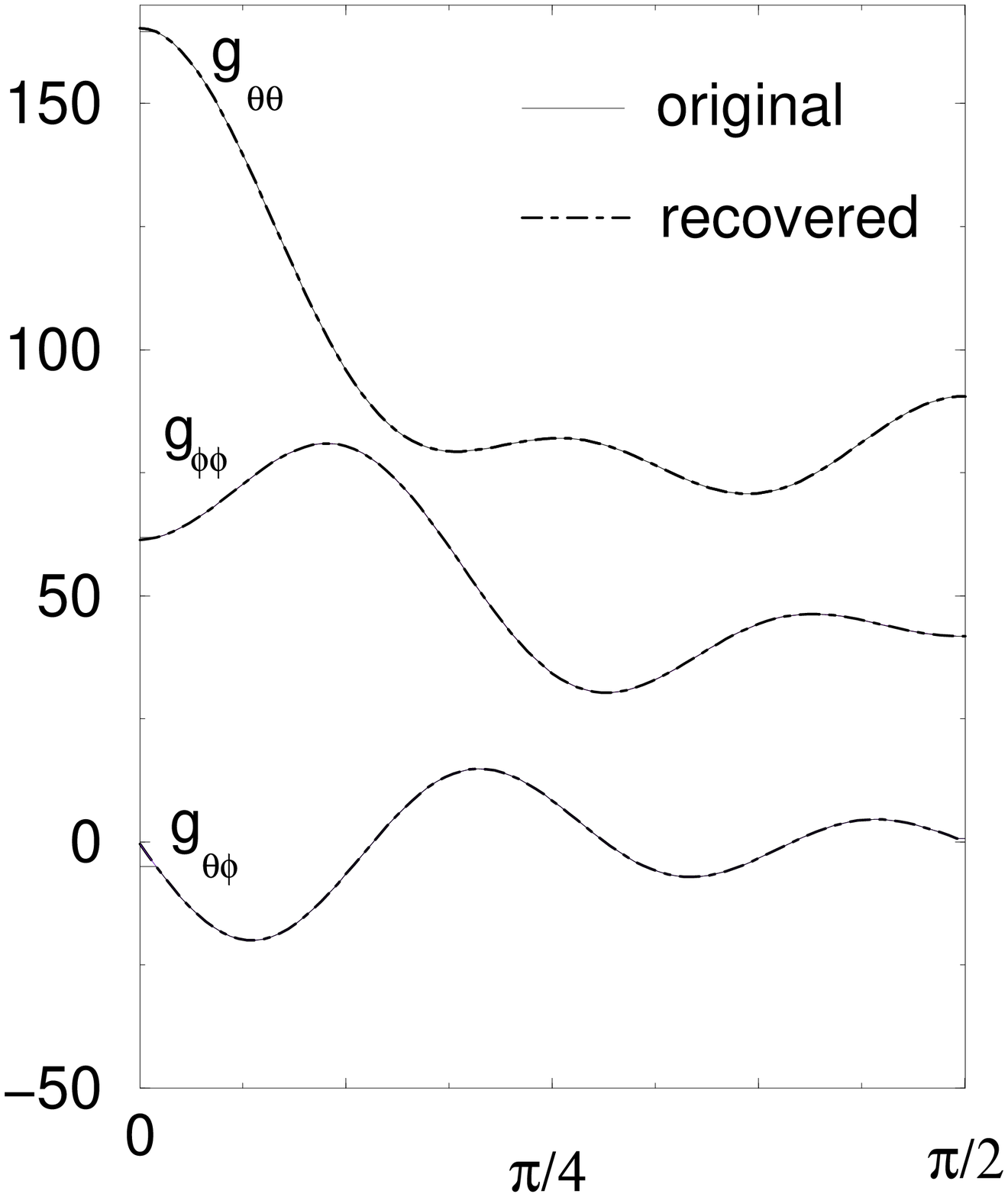}
\end{minipage}
\begin{minipage}{1.6in}
\epsfxsize=1.6in
\epsfysize=3.4in
\epsfbox{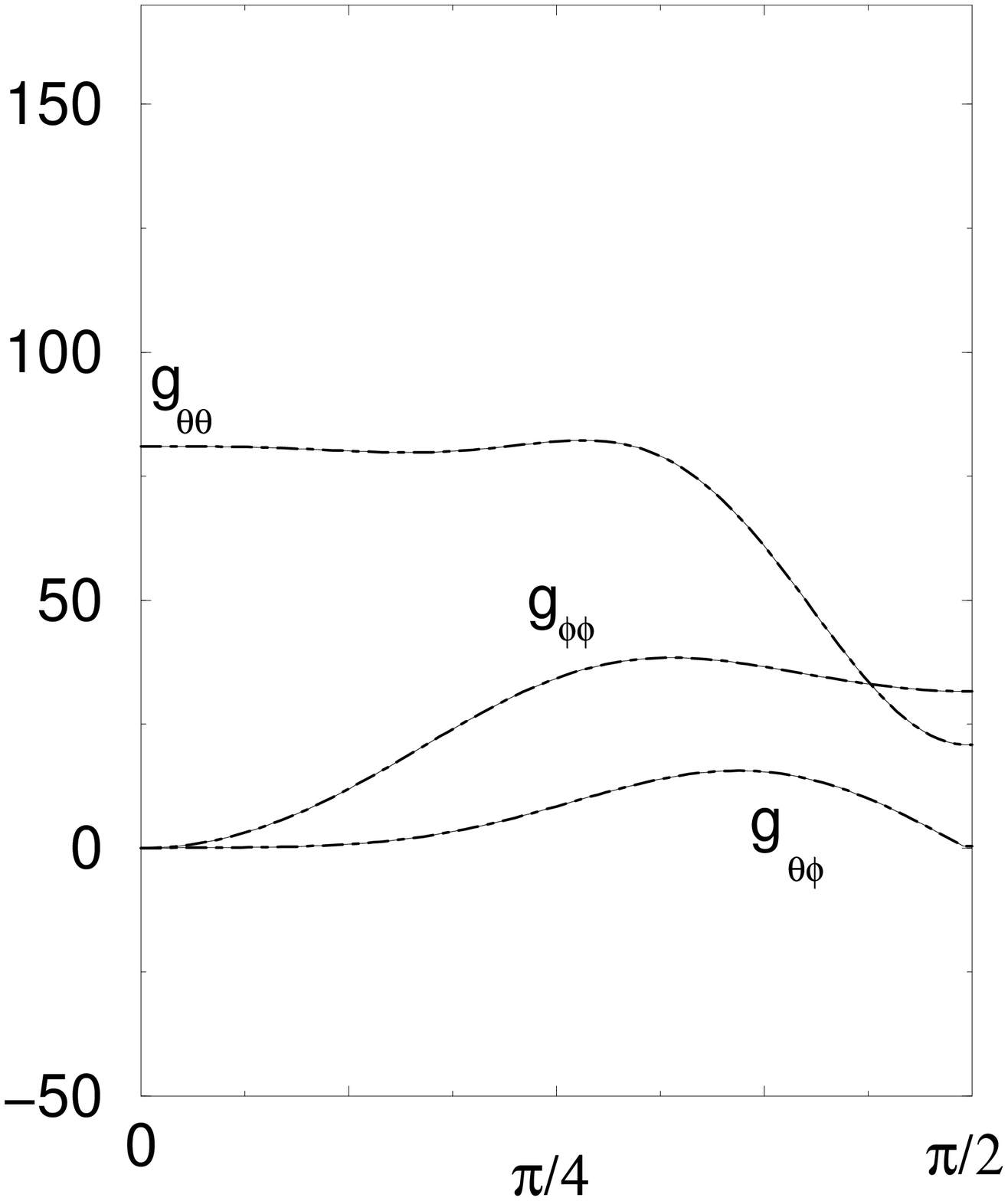}
\end{minipage}
\vspace{0.2in}
\caption{The difference in the metric components for the embedding of
the test surface described in the text. On the left panel we show the
line $\theta=\pi/4$ and on the right panel the line $\phi=\pi/4$.}
\label{fig:handmade_metric}
\end{figure}

Figure~\ref{fig:handmade_metric} shows a direct comparison of the
metric components $g_{\theta \theta}$, $g_{\theta \phi}$ and $g_{\phi
\phi}$ along the lines $\theta =\pi/4$ and $\phi=\pi/4$, i.e in the
middle of the computational domain.

The final value for the embedding function in this test was $F=1.6
\times 10^{-5}$, but we have found that we can easily decrease this
value by refining the numerical grid on the surface.


\subsection{An axisymmetric example: Rotating black holes}

As already mentioned in section~\ref{sec:intro}, a well-known set of
axisymmetric surfaces whose embeddings have been studied, first by
Smarr~\cite{Smarr73b}, and also as a test case in~\cite{Nollert98}, is
that of the horizons of rotating black holes.  In the static Kerr
case, the metric of the horizon is given by
\begin{equation}
d\sigma^2 = \rho^2 \, d\theta^2 + \frac{\sin^2\theta}{\rho^2} \,
\left( r^2 + a^2 \right) \,
d\phi^2 \, ,
\end{equation}
with
\begin{equation}
\rho^2 = r^2 + a^2 \cos^2\theta \, , \quad
r = m + \sqrt{m^2-a^2} \, ,
\end{equation}
and where $m$ and $a$ are two parameters representing the mass of the
black hole and its angular momentum respectively.

\begin{figure}
\epsfxsize=3.4in
\epsfysize=3.4in
\epsfbox{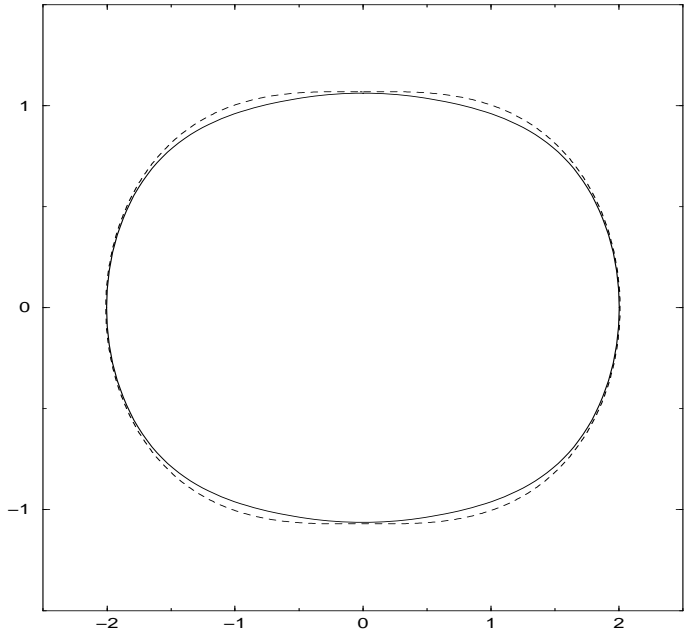}
\vspace{0.2in}
\caption{Embedding of a Kerr black hole with $a/m=\sqrt{3}/2$.  The
dotted line is the embedding we obtained using our minimization
algorithm. The solid line is an embedding of the same surface computed
with an axisymmetric algorithm.}
\label{fig:lastkerr}
\end{figure}

It is well known that the Kerr horizon is globally embeddable in flat
space only for $a/m \leq \sqrt{3}/2$~\cite{Smarr73b}.  Using our
embedding code, we have been able to successfully recover these
embeddings when they exist.  As an example we show in
Fig.~\ref{fig:lastkerr} the embedding obtained for the last
embeddable case $a/m=\sqrt{3}/2$.  The solid line shows the embedding
obtained with the axisymmetric algorithm described above, and the
dotted line the one obtained with our minimization algorithm.

\begin{figure}
\epsfxsize=3.4in
\epsfysize=3.4in
\epsfbox{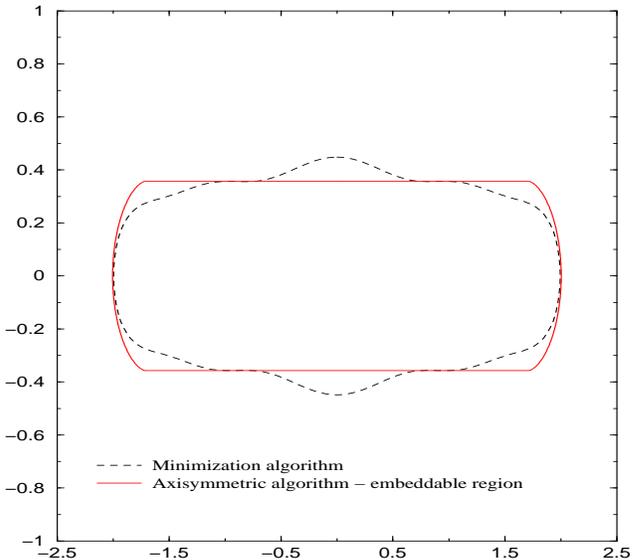}
\vspace{0.2in}
\caption{An attempt to embed a Kerr black hole horizon with
  $a/m=0.99$.  The dotted line is the output of the minimization
  algorithm. The solid line is an embedding of the same surface made
  with an axisymmetric algorithm.  The flat line on the top represents
  the region where an embedding in flat space does not exist.}
\label{fig:max_kerr}
\end{figure}

\begin{figure}
\epsfxsize=3.4in
\epsfysize=3.4in
\epsfbox{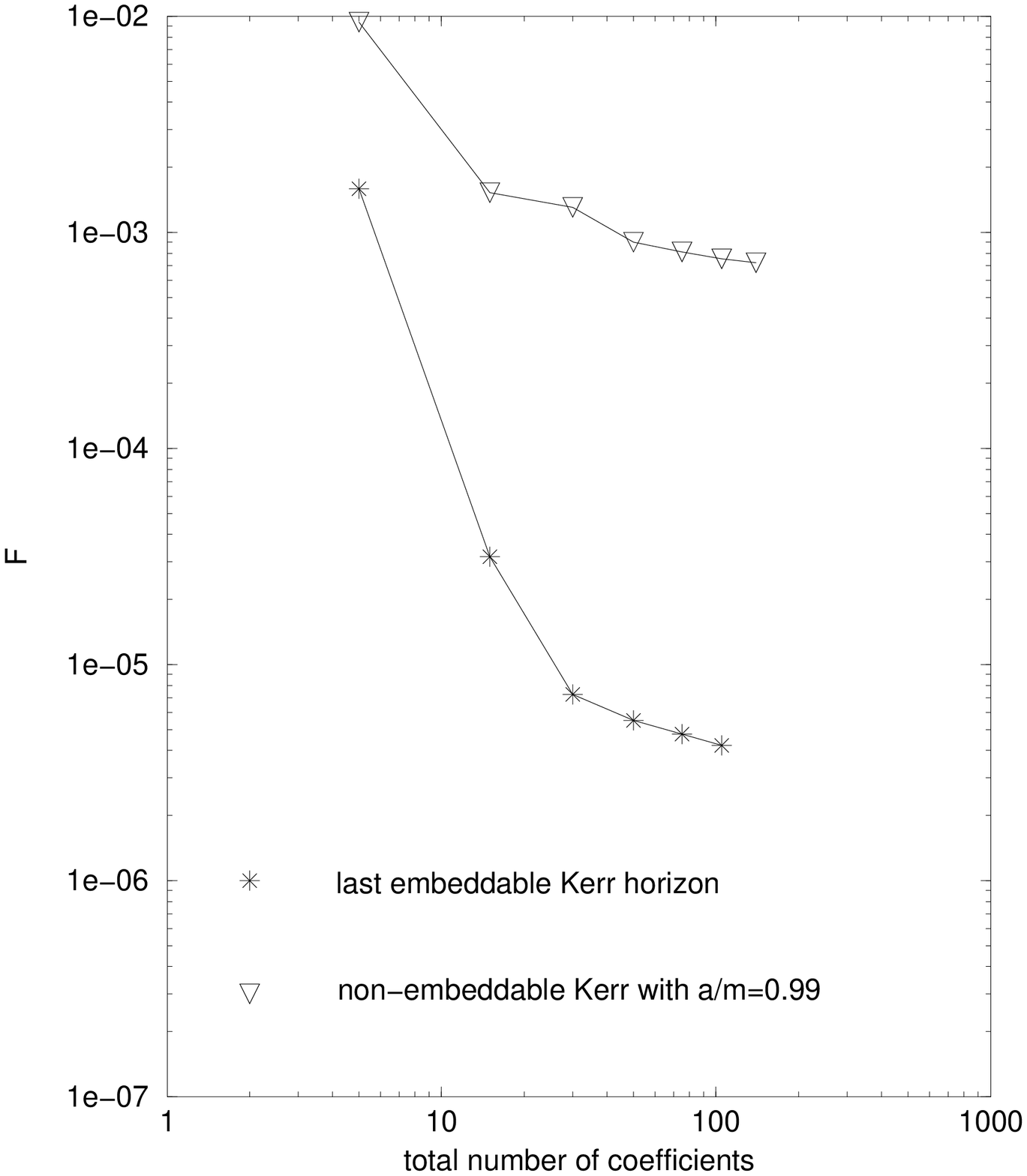}
\vspace{0.2in}
\caption{The value of the embedding function $F$ versus the total
number of coefficients for the two Kerr black holes discussed above.
The triangles correspond to $a/m=0.99$ and the stars to
$a/m=\sqrt{3}/{2}$.}
\label{fig:conv_kerr}
\end{figure}

When $a/m \geq \sqrt{3}/2$, a global embedding does not exist, and our
method fails as expected. Figure~\ref{fig:max_kerr} shows the result
of an attempt to embed a Kerr black hole horizon in the case when
$a/m=0.99$. The dotted line is the output of our minimization
algorithm and the solid line is an embedding of the same surface made
with an axisymmetric algorithm in the embeddable region, plus a flat
top in the region where the embedding does not exist. The axisymmetric
method is a {\em local} constructive method, and hence it is able to
build the embedding surface from one point to the next where it exists
(in this case starting from the equator).  Since our method is {\em
  global} we get the embedding wrong everywhere.  As currently
implemented, our method will insist on trying to find a global
embedding, and will settle on a shape that minimizes the function $F$.
If the embedding does not exist, the minimum value of $F$ found will
be clearly different from zero.  This is easily seen by examining the
residual function $F$ for the embeddings of the Kerr horizons.  In
Fig.~\ref{fig:conv_kerr} we show the value of the minimum of $F$ found
with our algorithm, plotted against the total number of expansion
coefficients, for the cases $a/m=\sqrt{3}/{2}$ (solid line) and
$a/m=0.99$ (dotted line).  One can see that for the non-embeddable
case, $F$ stops decreasing at a value that is more than 2 orders of
magnitude larger than the one we obtain when the embedding exists.

Our code might be adjusted in the future for finding partial
embeddings by reducing the integration domain in the definition of $F$,
Eq.~(\ref{eq:horizonfunction}), to something smaller than the full
sphere. This approach might lead to correct partial embeddings of
surfaces that cannot be embedded globally. The disadvantage would be
that one would have to guess the domain where the embedding exits
before starting the computation.


\subsection{Black hole plus Brill wave}

We now move to the case of numerically generated, distorted black holes.
Schwarzschild black holes distorted by Brill waves~\cite{Brill59} have
been extensively studied in numerical
relativity~\cite{Allen98a,Camarda97b,Anninos94c,Abrahams92a}.  The
axisymmetric data sets used for numerical evolutions consist of a
Schwarzschild black hole distorted by a toroidal, time-symmetric
gravitational wave.

It is convenient to describe the metric of the black hole plus Brill
wave spacetime in a spherical-polar like coordinate system
$(\eta,\theta,\phi)$ were $\eta$ is a logarithmic radial coordinate
defined by $\eta = \ln(2r/M)$ and $(\theta,\phi)$ are the standard
angular coordinates.  In these coordinates, the spatial metric has the
form~\cite{Bernstein93a,Anninos94c}
\begin{equation}
dl^2 = \Psi^4 \left[ e^{2q} \left( d\eta^2 + d\theta^2 \right)
+ r^2 \sin^2 \theta d\phi^2 \right] \, ,
\label{eq:BHplusBW}
\end{equation}
where both $q$ and $\Psi$ are functions of $\eta$ and $\theta$ only. In
order to satisfy the appropriate regularity and fall-off conditions,
the function $q$ has been chosen in the following way
\begin{equation}
q(\eta,\theta) = a \sin^n\theta \, \left[ e^{- \left( \frac{\eta+b}{\omega}
\right)^2} + e^{-\left( \frac{\eta-b}{\omega} \right)^2} \right] \, ,
\end{equation}
with $n$ is an arbitrary even number larger than $0$.  The parameter
$a$ characterizes the amplitude of the Brill wave, while the
parameters $b$ and $\omega$ characterize its radial location and width
respectively.  Having chosen the form of the function $q$, the
hamiltonian constraint is solved numerically for the conformal factor
$\Psi$. An isometry condition is imposed at a coordinate sphere to
guarantee that the final spacetime will contain a black hole.

Notice that the metric~(\ref{eq:BHplusBW}) is the three-dimensional
metric of space, and not the two-dimensional metric of the apparent
horizons.  The apparent horizons for these data sets have to be
located numerically.  Once these horizons are found, their
two-dimensional metric can be computed from the three-dimensional
metric given above, using the expressions given
in~(\ref{eq:2Dfrom3D}).

\begin{figure}
\epsfysize=3.4in
\epsfxsize=3.4in
\epsfbox{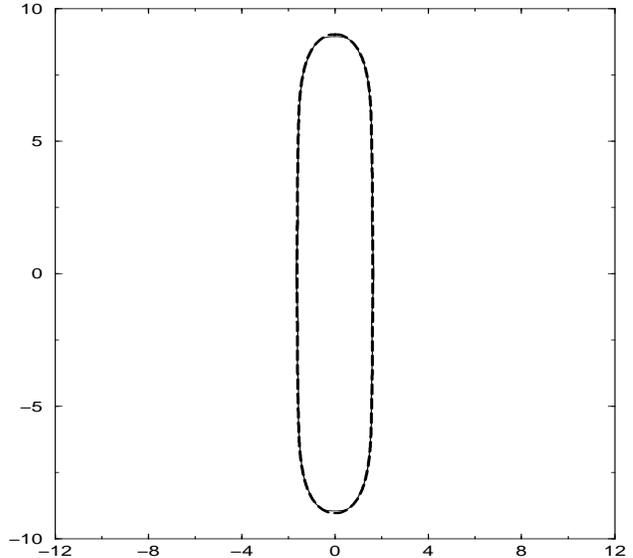}
\caption{Embedding of the apparent horizon of a black hole plus Brill
  wave data set corresponding to the parameters $(a=1.0, b=0.0,
  \omega=1.0, n=2)$.  The dotted line is the embedding obtained with
  our minimization algorithm and the solid line the embedding of the
  same surface obtained by Anninos et al.}
\label{fig:long}
\end{figure}

The horizons from these axisymmetric black hole plus Brill wave data
sets and their embeddings have been studied previously
in~\cite{Anninos94c} and we have been able to reproduce their results
using our algorithm.  An example of this can be seen in
Fig.~\ref{fig:long}, where we show the embedding of the horizon of a
black hole plus Brill wave data set corresponding to the parameters
\begin{equation}
a=1.0 \, , \quad b=0.0 \, , \quad w=1.0 \, , \quad n=2 \, .
\end{equation}

In the figure, the dotted line shows the embedding obtained with our
minimization algorithm, and the solid line shows the embedding of the
same surface obtained in~\cite{Anninos94c}.  Notice how the intrinsic
geometry of the horizon is far from spherical.

\subsection{Application to Full 3D Spacetimes}

Having tested our algorithm on both analytic and numerically generated
axisymmetric black hole spacetimes, we now turn to the case of full 3D
black hole spacetimes for which our method was developed.

The axisymmetric black hole plus Brill wave data sets
from~\cite{Anninos94c} have been generalized in~\cite{Camarda97a} to
full 3D by multiplying the Brill wave function $q$ by a factor that
has azimuthal dependence to obtain
\begin{eqnarray}
q(\eta,\theta,\phi) &=& a \sin^n\theta \, \left( 1 + c \, \cos^2 \phi \right)
\nonumber \\
&&  \left[ e^{- \left( \frac{\eta+b}{\omega} \right)^2)}
+ e^{-\left( \frac{\eta-b}{\omega} \right)^2} \right] \, ,
\end{eqnarray}
were $c$ is an arbitrary parameter characterizing the non-axisymmetry
of the Brill wave.

\begin{figure}
\epsfxsize=3.4in
\epsfysize=3.4in
\epsfbox{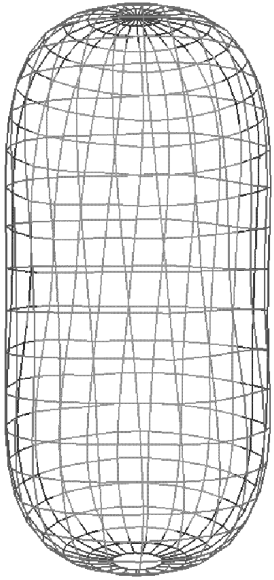}
\vspace{0.2in}
\caption{Embedding of the apparent horizon for the non axisymmetric
  black hole plus Brill wave data set corresponding to the parameters
  $(a=1.0, b=0.0, \omega=1.0, n=4, c=0.4)$. Although the metric has a
  non-trivial non-axisymmetric contribution, the surface looks quite
  axisymmetric.}
\label{fig:brill_nonaxi}
\end{figure}

We have computed embeddings of the non axisymmetric apparent horizons
obtained in this case.  Here we will show examples of two such
horizons. First we consider the embedding of the apparent horizon for
the data set with parameters
\begin{equation}
a=0.3 \, , \quad b=0.0 \, , \quad \omega=1.0 \, , \quad n=4 \, ,
\quad c=0.4 \, ,
\end{equation}
which corresponds to a relatively small non-axisymmetric distortion of
the black hole.  Figure~\ref{fig:brill_nonaxi} shows the embedding of
the corresponding apparent horizon.  Notice how the surface looks
quite axisymmetric even though we have added a non-trivial
non-axisymmetric contribution to the metric.  It is clear that the
non-axisymmetry of the metric components is to a large degree a
coordinate effect.  Numerical evolutions of such black holes do show
radiation in non-axisymmetric modes of gravitational radiation.
However the mass energy carried away by the non-axisymmetric modes is
much smaller than the energy of the axisymmetric
modes~\cite{Allen97a,Allen98a}. This is consistent with our result
showing that the horizon is almost axi-symmetric.

\begin{figure}
\begin{minipage}{1.6in} 
  \epsfxsize=1.6in
  \epsfysize=3.4in
  \epsfbox{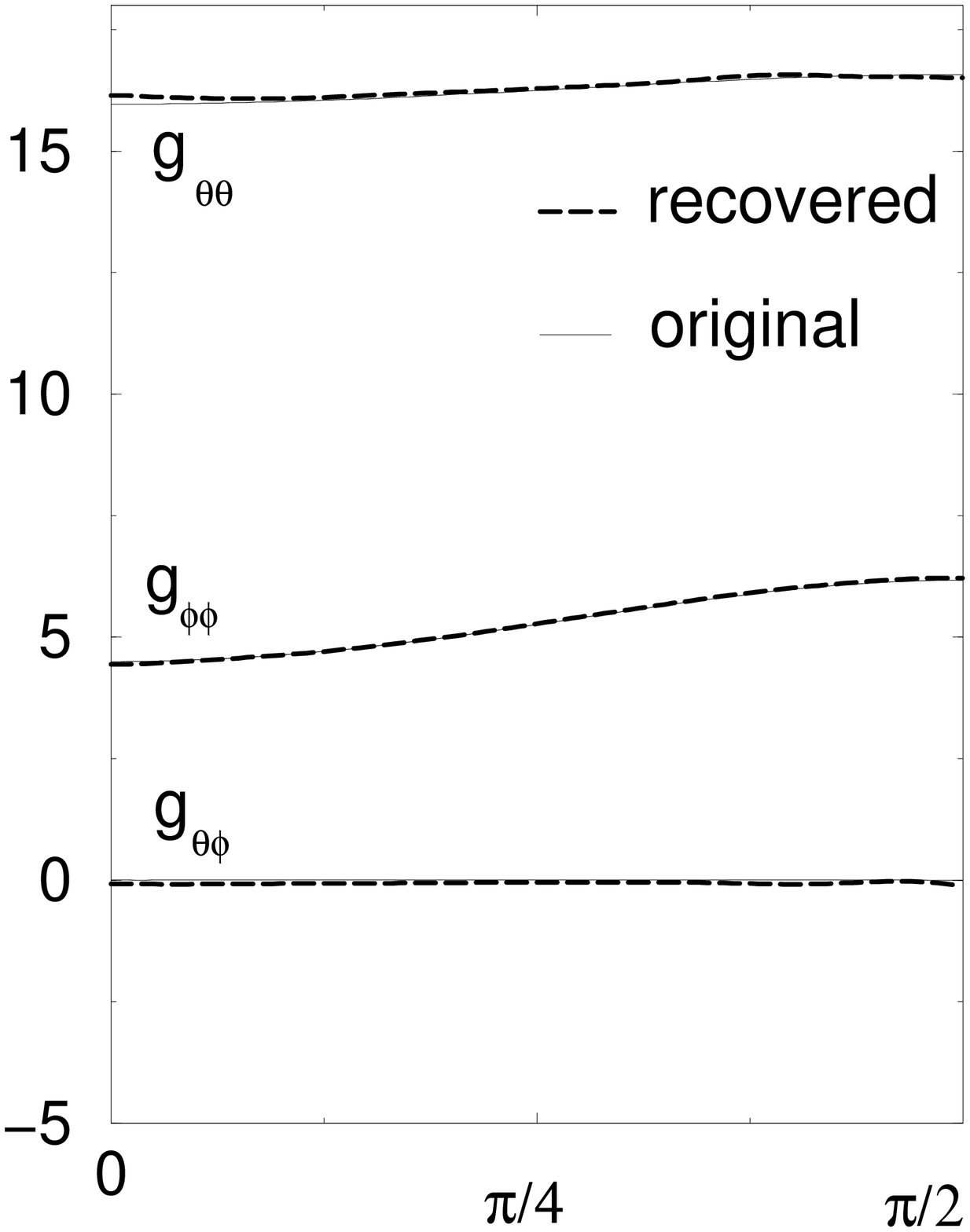}
\end{minipage}
\begin{minipage}{1.6in} 
  \epsfxsize=1.6in
  \epsfysize=3.4in
  \epsfbox{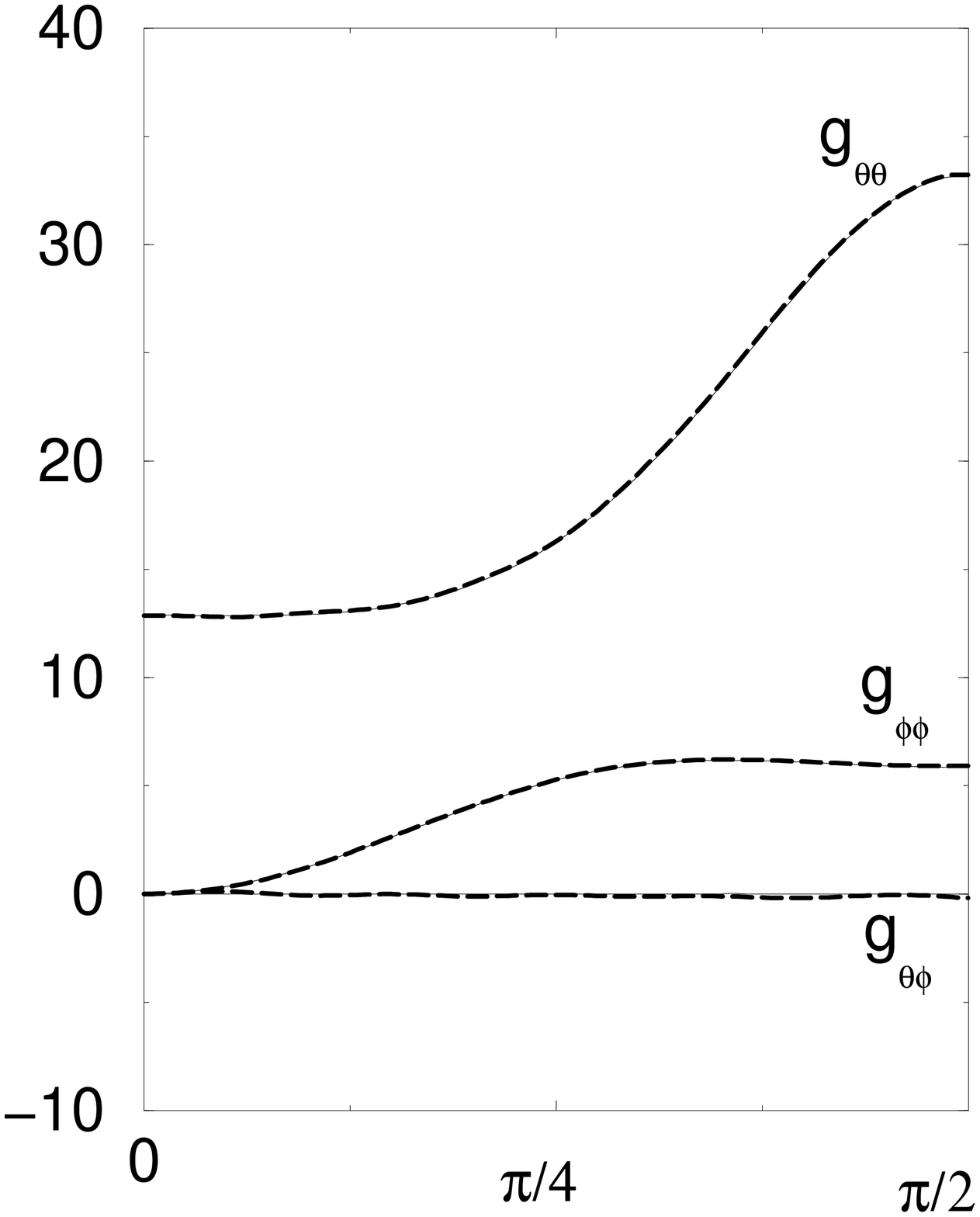}
\end{minipage}
\vspace{0.2in}
\caption{We show the angular metric components on the apparent horizon
and on the resulting embedding for the black hole plus Brill wave data
set corresponding to the parameters $(a=1.0, b=0.0, \omega=1.0, n=4,
c=0.4)$. On the left panel we show the line $\theta=\pi/4$ and on the
right panel the line $\phi=\pi/4$.}
\label{fig:fit_small_pertubation}
\end{figure}

Figure~\ref{fig:fit_small_pertubation} shows a direct comparison of
the different angular metric components on the apparent horizon and
the resulting embedding, along the $\phi = \pi/4$ and \mbox{$\theta =
  \pi/4$} lines.  We can see how the fit is very good in both cases.
Notice also how there is indeed some dependence of the metric components
on $\phi$.

\begin{figure}
\epsfxsize=3.4in
\epsfysize=3.4in
\epsfbox{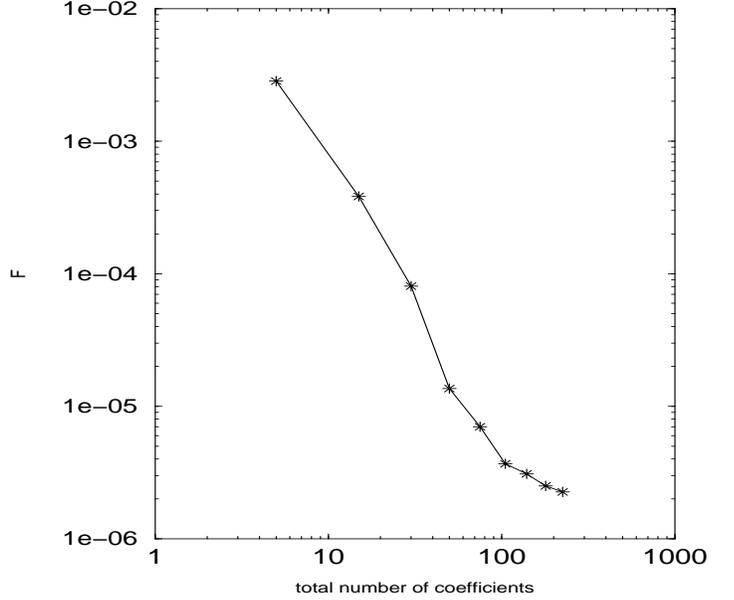}
\vspace{0.2in}
\caption{We show the value of the embedding function $F$ at the
  minimum in terms of the total number of expansion coefficients in a
  logarithmic scale for the black hole plus Brill wave data set
  corresponding to the parameters $(a=1.0, b=0.0, \omega=1.0, n=4,
  c=0.4)$.}
\label{fig:converg1}
\end{figure}

To check if our algorithm is converging to the correct embedding, we
show in Fig.~\ref{fig:converg1} the value of the embedding function
$F$ at the minimum, in terms of the total number of expansion
coefficients.  We clearly see that the value of $F$ is converging
exponentially to zero.

\begin{figure}
\begin{minipage}{3.4in} 
  \epsfxsize=3.4in
  \epsfysize=3.4in
  \epsfbox{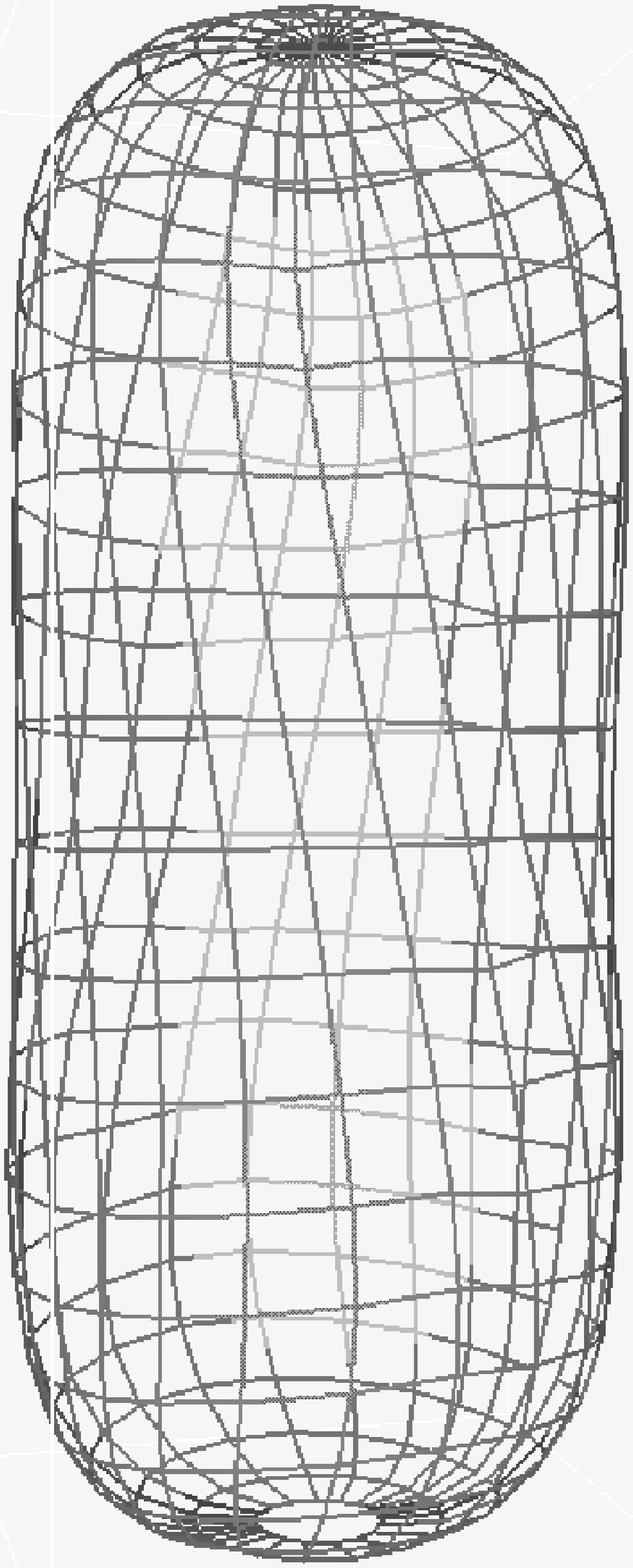}
\end{minipage}
\begin{minipage}{3.4in}
  \vspace{5mm}
  \epsfxsize=3.4in
  \epsfysize=3.4in
  \epsfbox{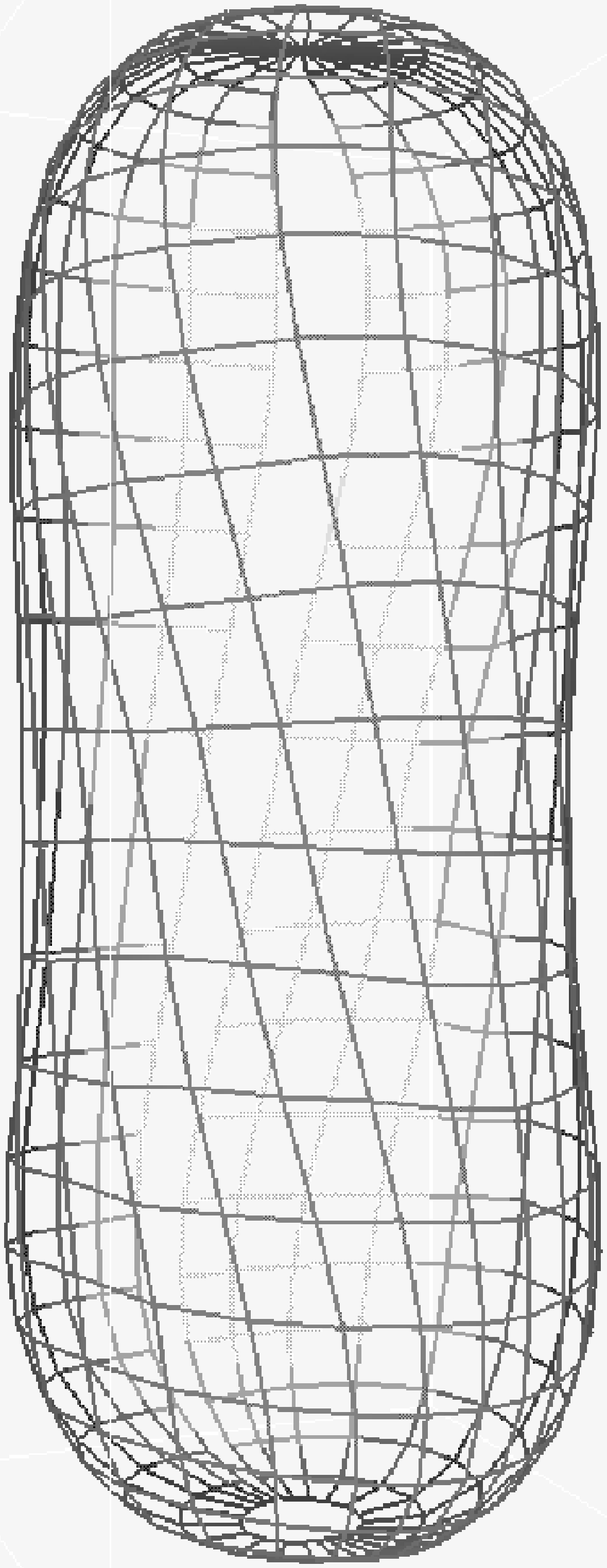}
\end{minipage}
\vspace{0.2in}
\caption{Two orientations of the embedding of the apparent horizon of
a black hole perturbed by a Brill wave with a higher non-axisymmetry
than in the previous example.}
\label{fig:large_perturbation}
\end{figure}

As a second example, we now consider the embedding of the apparent
horizon corresponding to the black hole plus Brill wave data set with
parameters
\begin{equation}
a=0.3 \, , \quad b=0.0 \, , \quad \omega=1.0 \, , \quad n=4 \, ,
\quad c=1.9 \, .
\end{equation}
In this case, the non-axisymmetry is considerably larger, and one can
see from Fig.~\ref{fig:large_perturbation} that the horizon
is clearly not axisymmetric.

\begin{figure}
\begin{minipage}{1.6in} 
  \epsfxsize=1.6in
  \epsfysize=3.4in
  \epsfbox{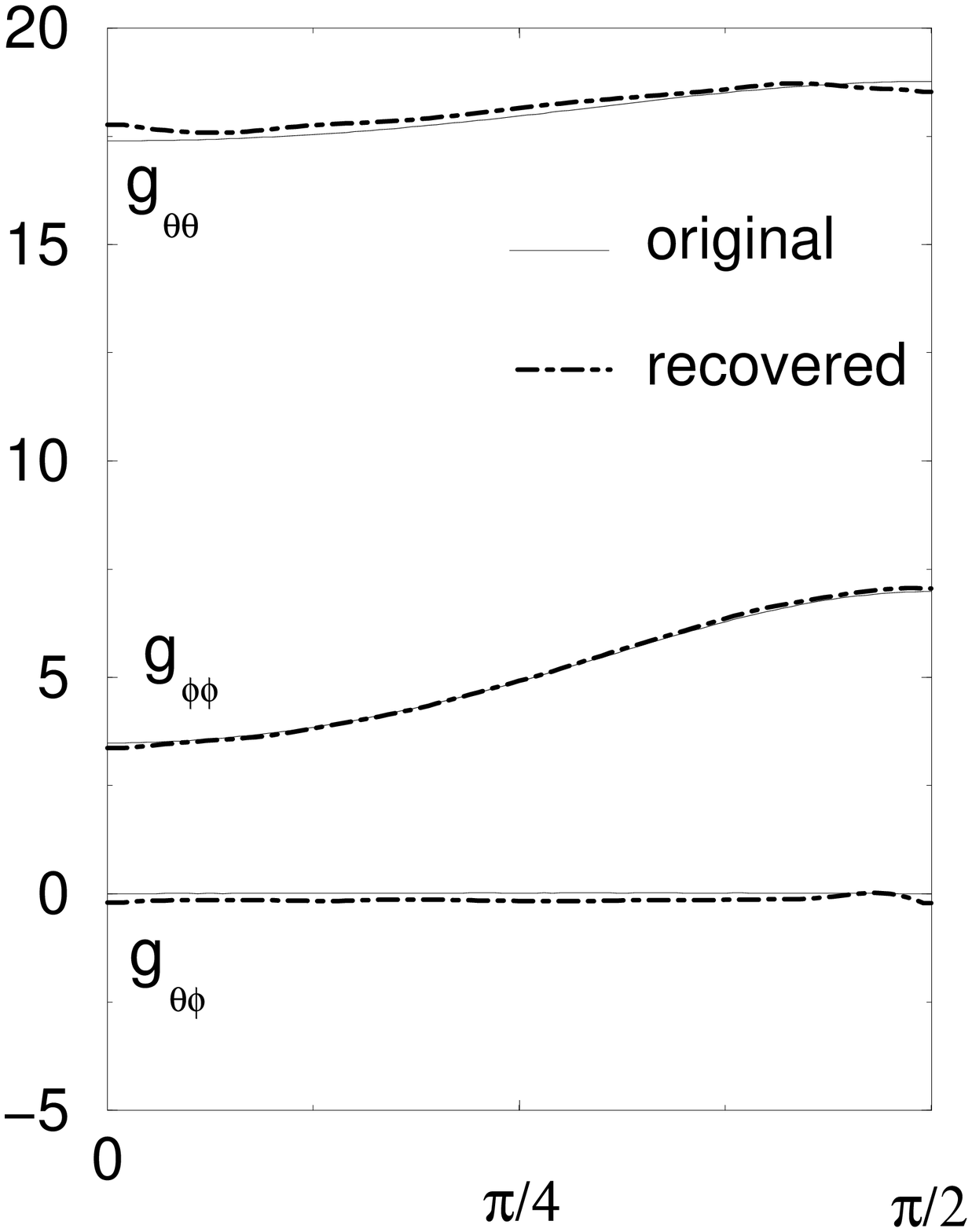}
\end{minipage}
\begin{minipage}{1.6in}
  \epsfxsize=1.6in
  \epsfysize=3.4in
  \epsfbox{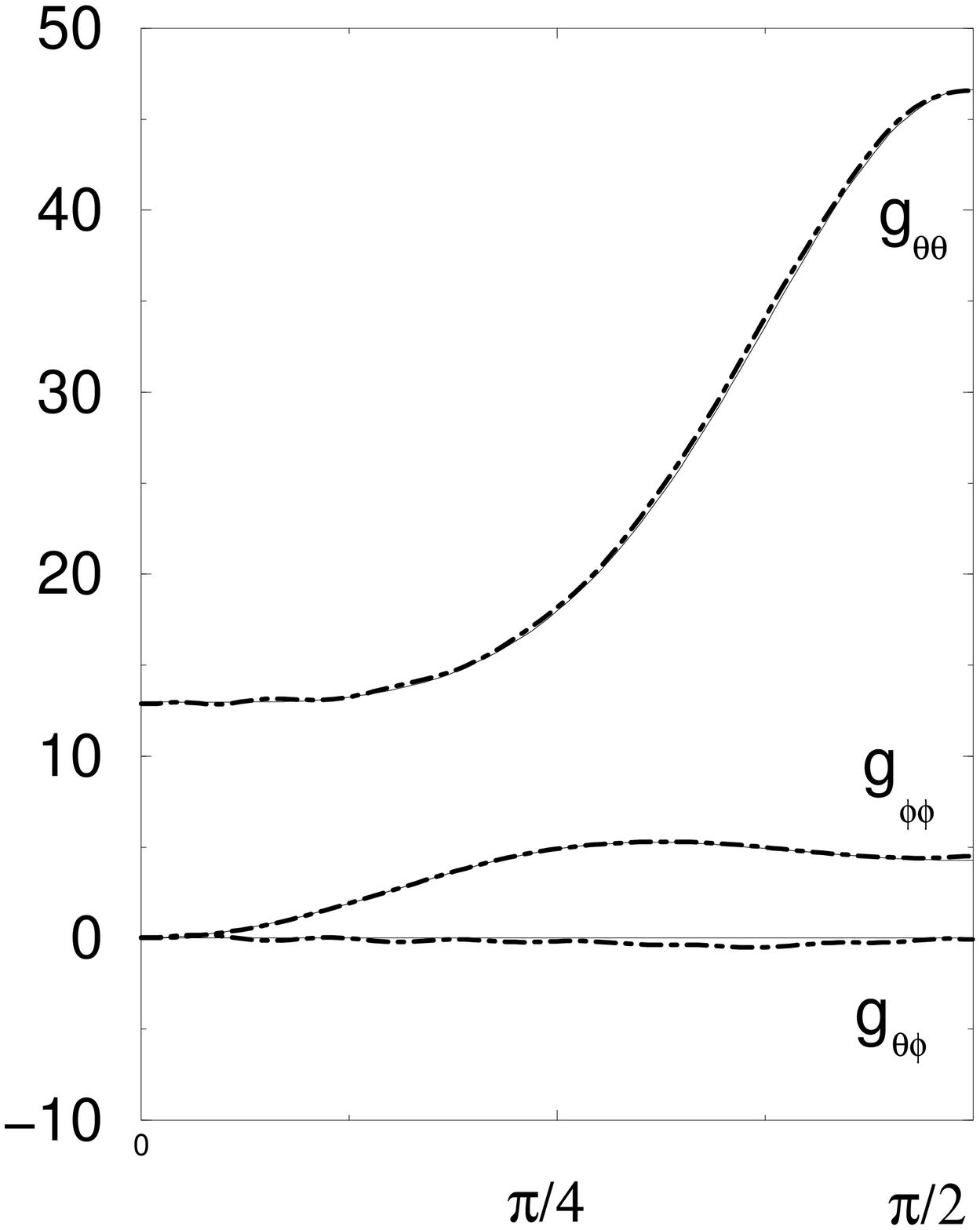}
\end{minipage}
\vspace{0.2in}
\caption{The three independent components of the metric for the
  black hole plus Brill wave will a larger non-axisymmetric
  perturbation.  On the left panel we show the line $\theta=\pi/4$
  and on the right panel the line $\phi=\pi/4$.}
\label{fig:fit_large_perturbation}
\end{figure}

In Fig.~\ref{fig:fit_large_perturbation} we show again a direct
comparison of the angular metric components on the apparent horizon
and the resulting embedding along the $\phi = \pi/4$ and $\theta =
\pi/4$ lines.  As before, the fit is very good.

\begin{figure}
\epsfxsize=3.4in
\epsfysize=3.4in
\epsfbox{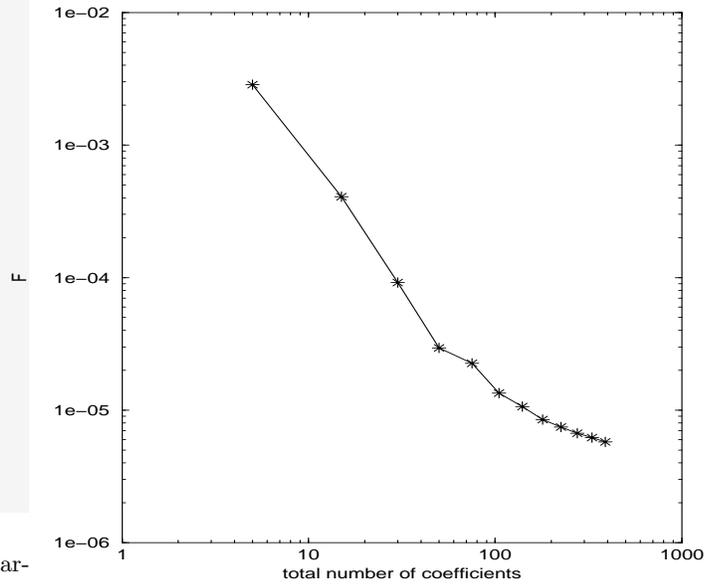}
\vspace{0.2in}
\caption{The value of the embedding function $F$ at the
  minimum in terms of the total number of expansion coefficients on a
  logarithmic scale for the black hole plus Brill wave with a larger
  non-axisymmetric contribution.}
\label{fig:converg2}
\end{figure}

Finally, in Fig.~\ref{fig:converg2} we show again the value of the
embedding function $F$ at the minimum in terms of the number of
expansion coefficients.  As before, the value of $F$ converges
exponentially to zero, but the convergence is slower than in the
previous example due to the higher degree of complexity of the
surface.

\section{Conclusions}
\label{sec:conclusions}

We have implemented and tested a new algorithm for computing isometric
embeddings of curved surfaces with spherical topology in flat space.
We define a function on the space of surfaces that has a global
minimum for the right embedding and we find this minimum using a
standard minimization algorithm.  In this paper we have discussed the
method and its applications to black hole visualization in numerical
relativity.  The method has been tested both on simple test surfaces
and on Kerr black hole horizons, and shown to correctly reproduce
known results within specified tolerances.  We have also used our
method construct embeddings of non-axisymmetric, distorted black holes
for the first time.  We observed that the non-axisymmetry of the
embedded surface is somehow smaller than one expects from just looking
at the metric.  This is consistent with the small amount of
gravitational radiation emitted in non-axisymmetric modes during the
numerical evolution of such systems~\cite{Allen98a,Allen97a}.

Our method is rather robust, and by construction produces smooth
surfaces, therefore avoiding some of the problems of previous methods.
One disadvantage of our method is that the expansion in spherical
harmonics implies that it can only be used to embed ray-body surfaces
(i.e. surfaces such that any ray coming from the origin intersects the
surface at only one point), still we expect most black hole horizons
to have this property.  The main disadvantage, however, is that the
method is very time consuming due to the fact that minimization
algorithms in general are slow.  Another problem is the fact that
minimization algorithms can easily get trapped in local minima.  In
order to avoid this we have found it necessary to increase the number
of coefficients one by one and to use at each step the result of the
previous step as initial guess, adding to the total amount of time the
algorithm needs to find the embedding.  As presently implemented, it
also cannot find partial embeddings when no global embeddings exist,
but straightforward modifications to the algorithm should permit this
in certain cases.

In the future, we will apply this method to study the dynamics of 3D
black hole horizons as a tool to aid in understanding the physics of
such systems.  Although our method has been applied in this paper only
to apparent horizons, it can clearly be applied to obtain embeddings
of event horizons as well, once they have been located in numerical
evolutions.

Our embedding algorithm has been implemented as a thorn in the Cactus
code~\cite{Cactusweb}, and is available for the community upon request
from the authors.


\acknowledgements

The authors would like to thank C.~Cutler, D.~Vulcanov and A.~Rendall
for many helpful discussions.  We are also grateful to R.~Takahashi
for suppling some of the horizon data used in the embeddings and to
P.~Diener for speeding up our code by changing the way in which we
computed Legendre polynomials.  We are specially thankful to W.~Benger
for modifying the visualization package Amira so that it can be used
for the three-dimensional visualization of embeddings.


\bibliographystyle{prsty}
\bibliography{bibtex/references}

\end{document}